# Current comparator for both AC and DC ratio measurements with $10^{-8}$-level accuracy

Hidekazu Muramatsu[1,2,6], Yuta Kainuma[1,2], Hiromitsu Kato[2,3], Norihiko Sakamoto[1], Tatsuji Yamada[1], Chiharu Urano[1,2], Hiroshi Abe[4], Shinobu Onoda[4], Takeshi Ohshima[4,5], Yuji Hatano[6], Mutsuko Hatano[6], Nobu-Hisa Kaneko[1,2], Yasutaka Amagai[1,2], and Takayuki Iwasaki[6]

[1]National Metrology Institute of Japan (NMIJ), National Institute of Advanced Industrial Science and Technology (AIST)
[2]Global Research and Development Center for Business by Quantum-AI Technology (G-QuAT), National Institute of Advanced Industrial Science and Technology (AIST)
[3]Advance Power Electronics Research Center, National Institute of Advanced Industrial Science and Technology (AIST)
[4]National Institutes for Quantum Science and Technology (QST)
[5]Department of Material Science, Tohoku University
[6]Department of Electrical and Electronic Engineering, School of Engineering, Institute of Science Tokyo
Corresponding authors: T. Iwasaki (iwasaki.t.c5b4@m.isct.ac.jp), Y. Amagai (y-amagai@aist.go.jp), and H. Muramatsu (muramatsu.h@aist.go.jp)

**Accurate measurements of alternating current (AC) and direct current (DC) ratios are fundamental to electric power metrology. However, conventional current comparators for AC and DC typically rely on distinct technologies—electromagnetic induction for AC and superconducting quantum interference devices for DC. This technological divide leads to a fragmented and complex traceability system. Bridging this gap is critical for developing unified current standards that meet the demands of emerging power technologies. In this work, we present a compact, room-temperature AC/DC current comparator that integrates a diamond-based magnetometer using nitrogen-vacancy centers. The device achieves an accuracy of $10^{-8}$ for both AC and DC signals and supports a system bandwidth up to 300 Hz, without the need for cryogenics. It surpasses the performance of typical AC comparators, offering ten-fold higher accuracy, and matches that of state-of-the-art DC comparators. This unified, cryogenics-free solution not only enhances precision and versatility but also expands the applicability of the system to DC resistance bridges in quantum electrical standards.**

## Introduction

Distributed energy sources such as solar and wind, along with power electronic devices, are increasingly being integrated into modern power grids. These advancements introduce not only conventional power-line frequency currents (50 Hz/60 Hz) but also harmonic and direct currents (DC). Given that current transformers are widely used to scale high currents to levels suitable for measurement and protection in industrial settings, accurate current ratio measurements across a broad frequency range are essential for reliable grid monitoring.

Traditionally, current ratio measurements rely on current comparators[1-3], which determine the ratio between two currents by detecting the null magnetic flux in a high-permeability core. Conventional alternating current comparators[3,4] (ACCs), which use detection windings based on electromagnetic induction, serve as primary standards in several national metrology institutes (NMIs), achieving accuracies on the order of $10^{-6}$. While this level of accuracy suffices for many applications, further improvements are limited by the core's physical properties and the inherent constraints of electromagnetic detection. For DC applications, cryogenic current comparators[2,5,6] (CCCs) or DC comparators[7-11] (DCCs), which use superconducting quantum interference devices[12-14] (SQUIDs) or fluxgate sensors, can reach accuracies of approximately $10^{-9}$.

However, a significant limitation of existing metrology systems lies in their fragmentation: AC and DC measurements require distinct comparator technologies. This separation complicates the calibration infrastructure and reduces measurement flexibility. Furthermore, conventional current comparators often involve bulky windings—up to 500 turns—and, in some cases, the need for cryogenic cooling. A single, compact device capable of performing accurate current ratio measurements across both AC and DC regimes without cryogenics would streamline traceability and broaden the applicability of high-precision current metrology in industrial environments.

To overcome these challenges, we leverage nitrogen-vacancy (NV) centers in diamond[15–18], which have emerged as powerful candidates for quantum magnetic-field sensing. NV centers allow magnetic field detection across a wide frequency range—from DC up to several MHz[19–21]—through optical interrogation of their electron spin states. They offer a broad dynamic range near zero magnetic field[22], high sensitivity without the need for cryogenic cooling, and resilience to temperature fluctuations, making them ideally suited as null detectors for both AC and DC current ratio measurements. While NV-based magnetometers have previously been used for magnetic field sensing, their use in precision current ratio metrology—particularly as unified AC/DC current comparators—has not yet been demonstrated.

In this study, we developed a compact current comparator integrating a diamond quantum sensor and successfully demonstrated accurate current ratio measurements at the $10^{-8}$ level across both AC and DC regimes. To achieve this goal, a low-noise and highly sensitive NV-diamond magnetometer was developed and embedded in the air gap of a magnetic core, eliminating the need for large detection windings and massive magnetic cores while improving the sensitivity. This work represents a significant step toward unifying AC and DC current ratio measurements in a room-temperature platform, thereby simplifying electrical metrology traceability and expanding the applicability of quantum sensors in precision instrumentation.

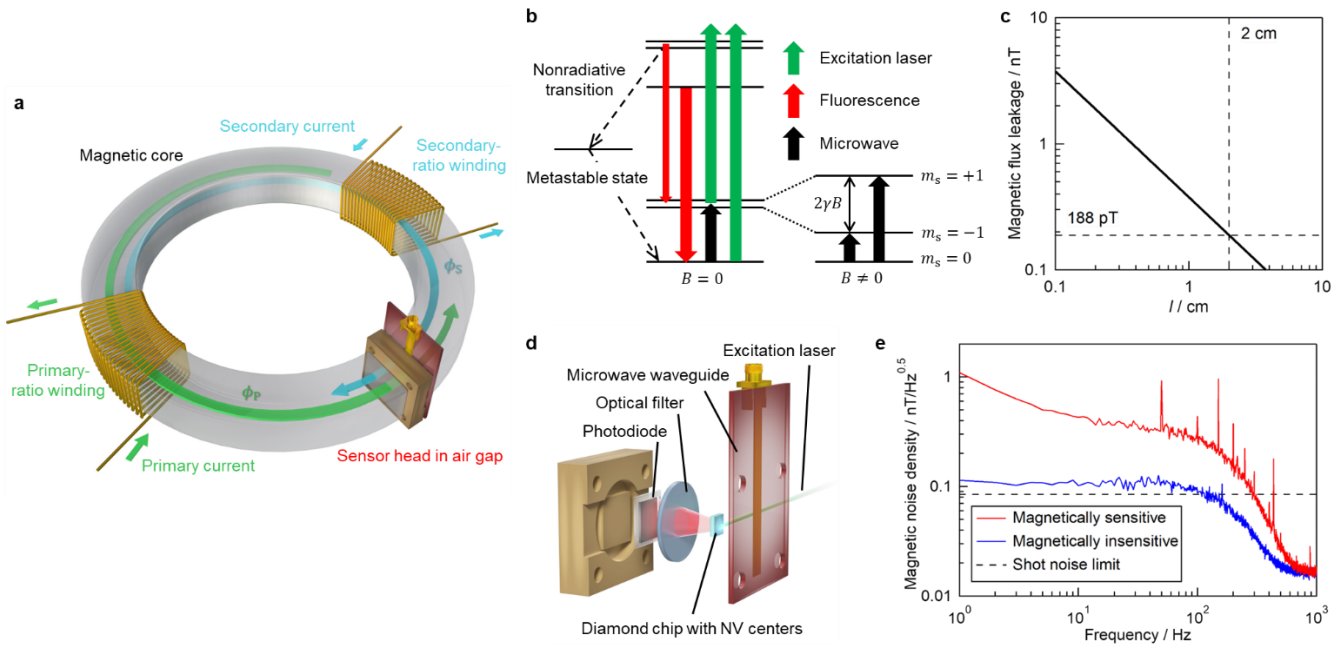

**Figure 1 | Precision current ratio measurement using a solid-state quantum sensor. a,** Schematic of an NV-diamond-based current comparator. The primary- and secondary-ratio windings are wound around the air-gapped magnetic core. Although the figure depicts the two windings as being concentrated on diametrically opposite localized regions of the core, they are in fact uniformly distributed over the entire magnetic core. The diamond quantum sensor placed in the air gap of the core measures the residual magnetic flux produced by the currents in the ratio windings. $\phi_P$ and $\phi_S$ are magnetic fluxes generated by the currents. **b,** Energy level diagram of the NV center. The $m_s = \pm 1$ ground states are split by an external magnetic field with the width proportional to the magnitude of the magnetic field $B$ and gyromagnetic ratio $\gamma$. **c,** Calculated magnetic flux in the air gap as a function of the air-gap length. **d,** Schematic of the sensor head of the diamond magnetometer. The sensor head includes a (111)-oriented diamond chip placed on a microwave antenna, an optical filter, and a photodiode within a width of approximately 1 cm. **e,** Magnetic noise spectrum of the present system. The red solid line shows the magnetic noise spectral density under magnetically sensitive conditions. The blue solid line shows the magnetic noise spectrum excited by microwaves at non-resonant frequencies. The dotted line represents the calculated photon shot-noise limit.

## Design of a current comparator measuring both AC and DC current ratios

As part of our effort to develop a current comparator capable of accurately measuring both AC and DC current ratios at a level suitable for primary standards, we began by analyzing the structure and operating principle of conventional devices. A traditional alternating current comparator (ACC) determines the ratio between two currents—applied to the primary and secondary windings—wound around a high-permeability toroidal magnetic core (without an air gap). These currents generate opposing magnetic fluxes within the core, which ideally cancel each other out. The balance point, or null condition, is detected by a sensing winding that measures the residual magnetic flux in the core. Typically, such magnetic cores have a diameter of around 30 cm, and the detection winding may consist of up to 500 turns, resulting in a total weight of approximately 20 kg.

A nickel-iron-based alloy (permalloy) is commonly used as the magnetic core material due to its high magnetic permeability and low coercivity, making it ideal for precision magnetic applications. Following the configuration of a conventional ACC, we designed our current comparator to operate at the most fundamental current ratio—namely, a one-to-one ratio. The developed one-to-one current comparator consists of an air-gapped magnetic core integrated with a low-noise diamond magnetometer, along with primary and secondary windings (Fig. 1a). The residual magnetic flux in the air gap is measured by the diamond magnetometer, which incorporates an ensemble of nitrogen-vacancy (NV) centers in diamond.

We employed continuous-wave optically detected magnetic resonance (cw-ODMR) for magnetic field detection. In cw-ODMR, the energy splitting between the $m_s = \pm 1$ spin states (Fig. 1b), where $m_s$ is the magnetic quantum number, is inferred from the fluorescence intensity. The NV spin states are optically initialized using a continuous-wave green laser and manipulated by applying microwaves at their resonant frequencies. The Zeeman-induced splitting is proportional to the magnitude of the magnetic field component along the NV axis, allowing precise determination of the residual magnetic flux from the measured resonance frequencies.

To assess the sensitivity requirements for the diamond magnetometer, we calculated the expected magnetic flux amplitude within the core using a magnetic equivalent circuit model for a core with a 2 cm air gap.

The analysis revealed that the magnetic flux, which depends on factors such as relative permeability, cross-section area of the magnetic core, and the length of the magnetic path of the core, varies with the air-gap length (Fig. 1c). Assuming a typical current measurement resolution of $1 \times 10^{-6}$ A and using three turns—common in ACCs employed by NMIs—our calculations indicate that the diamond magnetometer must detect a residual magnetic flux of approximately 188 pT in the 2 cm air gap at a current amplitude of 1 A.

A schematic of the sensor head of the NV-based magnetometer is shown in Fig. 1d. The design integrates a (111)-oriented diamond directly onto a microwave antenna, allowing compact placement within the air gap of the magnetic core without requiring external optics or bulky components. The sensor head consists of a (111)-oriented diamond chip containing an ensemble of NV centers, positioned on a microwave antenna for spin manipulation. An optical filter is used to block background light, including the excitation laser, and a photodiode detects the NV red fluorescence. The diamond's (111) surface is oriented perpendicular to the direction of the residual magnetic flux in the air gap. The excitation laser is introduced from one of the side faces of the diamond chip. All components are integrated within a width of approximately 1 cm, enabling the sensor to fit inside the 2 cm air gap[23]. The magnetic core has an outer diameter of 10 cm, an inner diameter of 6 cm, and a thickness of 2 cm—roughly one-third the size of typical magnetic cores used in conventional ACCs. Its weight is approximately 1 kg, which is one-twentieth that of magnetic cores typically used in conventional ACCs.

To evaluate whether the diamond magnetometer can satisfy the requirement of 188 pT, we measured the noise spectra under both magnetically sensitive and magnetically insensitive conditions (Fig. 1e). While no current was applied to the primary and secondary ratio windings, a DC current was supplied to an auxiliary winding to generate an offset DC magnetic field. This offset was necessary to avoid the nonlinear response region near zero magnetic field in the ODMR spectrum. To measure the magnetic field magnitude, we employed a time-domain multiplexed frequency modulation technique[24], which enables simultaneous tracking of the two Zeeman-split resonance frequencies.

As shown by the blue solid line in Fig. 1e, setting the system to a non-resonant microwave frequency of 2 GHz resulted in a magnetically insensitive noise floor of approximately 100 pT/Hz$^{0.5}$ in the 10 Hz to 100 Hz frequency range. Under these magnetically insensitive conditions, the noise floor was primarily limited by photon shot noise, estimated to be 85 pT/Hz$^{0.5}$. In contrast, the red solid line shows the magnetic noise spectrum under magnetically sensitive conditions, where the noise floor increased to approximately 300 pT/Hz$^{0.5}$ at the fundamental power-line frequencies.

Since this 300 pT/Hz$^{0.5}$ noise floor is dominated by random noise sources, the magnetic sensitivity of the system improves with longer integration times, scaling inversely with the square root of the integration time.

Under this assumption, the target detection limit of 188 pT—corresponding to a current resolution of 1 μA in a 2 cm air gap—can be achieved by integrating for approximately 2.5 s, being comparable with a standard measurement time in electrical standards. This suggests that the developed diamond magnetometer is capable of supporting AC current ratio measurements with a resolution comparable to the $10^{-6}$ level of accuracy typically attained by conventional current comparators used as national standard at NMIs.

**Evaluation of Ratio Error and Stability**

We performed AC current ratio measurements to evaluate the performance of the developed current comparator. For this purpose, the primary and secondary ratio windings, each with 10 turns, were connected in series. In an ideal current comparator, the current ratio is solely determined by the turn ratio of the windings. However, practical deviations—due to leakage flux, stray capacitance, and other parasitic effects—introduce what is known as ratio error[1].

To quantify this ratio error, the residual magnetic flux in the air gap was measured using the NV-diamond magnetometer. A constant DC magnetic field offset was applied to position the NV resonance within its linear response regime. An AC current was then supplied to the series-connected ratio windings, generating a magnetic flux detectable by the sensor. The resulting time-domain magnetic field signal—comprising the AC component, noise, and DC offset—was analyzed using a discrete Fourier transform (DFT) to extract the flux component at the excitation frequency.

To assess performance at a representative power-line frequency, we measured the amplitude spectrum at a fixed input current of 1 A (Fig. 2a). The excitation frequency was set to 67 Hz to avoid interference from common 50 Hz/60 Hz power-line noise and its harmonics, which can degrade low-frequency magnetic measurements. The equivalent current difference was determined using a conversion coefficient relating the residual magnetic flux to current deviation at 67 Hz. As shown in the amplitude spectrum, a clear peak was observed at the expected frequency of 67 Hz. The measured noise floor of the current comparator system was approximately 1 μA, denoted by the dotted line, as shown in Fig. 2a. To investigate how the measured current difference—corresponding to the current ratio—varies with the frequency of the input current, we conducted AC current difference measurements across several frequencies. The measured current difference as a function of input frequency at a fixed current amplitude of 1 A was evaluated over a range of frequencies (Fig. 2b). The system exhibited a bandwidth of 300 Hz, which was limited by the low-pass filter setting of the lock-in amplifier. The observed frequency response may result from eddy current losses and hysteresis effects in the high-permeability magnetic core material[25].

Next, we examined whether the residual magnetic flux—and thus the current ratio error—is linearly proportional to the input current amplitude. To do so, we acquired magnetic field spectra while systematically varying the current amplitude. Both the measured current difference and corresponding residual flux scaled linearly with input current (Fig. 2c), confirming the system's stable and linear performance over a range of current

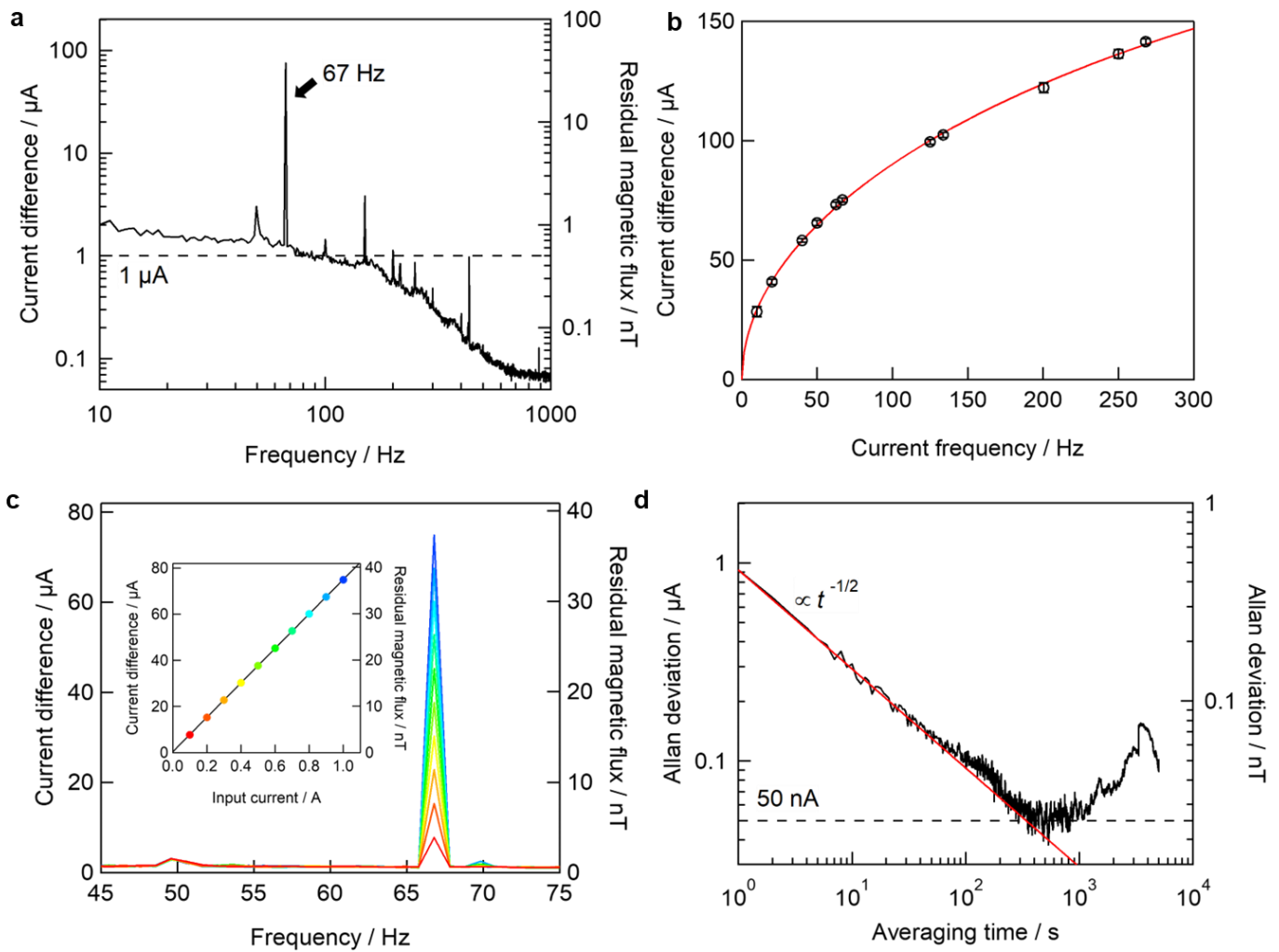

**Figure 2 | The performance of the developed current comparator for AC ratio measurements. a,** Amplitude spectra of the developed one-to-one current comparator obtained by a one-second measurement at a current frequency of 67 Hz. The left and right axes represent the equivalent current difference and residual magnetic flux, respectively. A clear peak appears at the expected frequency of 67 Hz. The noise floor of the present system is approximately 1 µA (denoted by the dotted line in **a**). **b,** Equivalent current difference as a function of frequency. The red solid line represents the fitting curve based on the model of the hysteresis loss and the eddy current loss. **c,** The equivalent current difference measured by the diamond quantum sensor is proportional to the amplitude of the input current, as shown in the inset of **c**. Each dot in the inset corresponds to the peak value of the same color spectrum. Each residual flux was averaged 100 times, and the error bars, which represent the standard deviation of the mean, were smaller than each marker. The slope of the solid line represents the ratio error of the current comparator at 67 Hz. **d,** Allan deviation at an amplitude of 1 A and a frequency of 67 Hz. The red solid line represents the theoretical curve assuming white noise. The horizontal dotted line denotes the detectable current limit, corresponding to 50 nA.

amplitudes.

Under typical operating conditions, we quantified the ratio error in the current comparator. At an input amplitude of 1 A and 67 Hz, the measured current difference was 76 µA, yielding a ratio error of 76 µA/A. Although this does not meet the stringent 1 µA/A requirement typically expected in metrological applications, the observed error reflects performance under the present experimental conditions and does not represent the fundamental limit of the developed system. Given the stability of the error over time and varying conditions, it is possible to correct for it through calibration, enabling the comparator to function as a traceable measurement instrument in practical applications.

**Statistical Analysis in AC ratio measurements**

To further evaluate the uncertainty associated with the ratio error, we analyzed the time dependence of the signal-to-noise ratio (SNR) using Allan deviation. The Allan deviation of the mean was calculated for various averaging times $t$ (Fig. 2d) to assess the statistical convergence behavior and the attainable precision limits of the current measurements. A clear trend of decreasing Allan deviation with increasing averaging time was observed, reaching a minimum value of 25 pT before increasing again. The observed $t^{-1/2}$ time dependence confirms that the noise in the system is dominated by uncorrelated random noise.

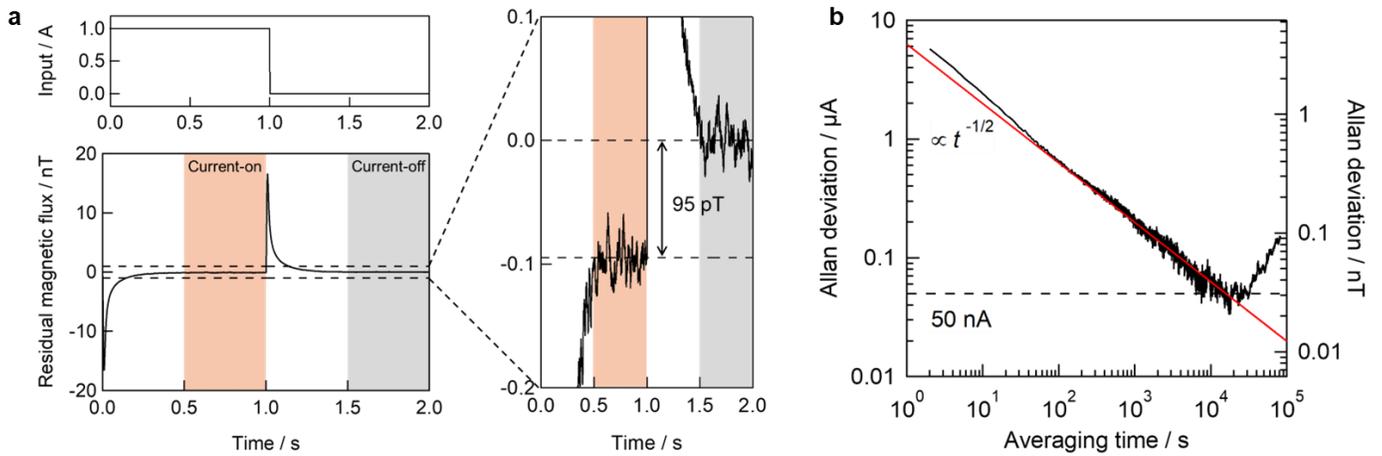

**Figure 3 | The performance of the developed current comparator for DC ratio measurements. a,** Time-domain signal of the residual magnetic flux. A square signal with a width of 1 s was applied to the ratio windings, and the residual magnetic flux was calculated as the difference between the signals in the current-on and current-off states except for the transient parts. The right figure shows an enlarged view of the near zero nanotesla region of the left figure. The signal of the residual flux was shifted so that the average value in the current-off state was zero. **b,** Allan deviation at a current of 1 A. The red solid line represents the theoretical curve assuming white noise. The horizontal dotted line denotes the detectable current limit, corresponding to 50 nA.

At 67 Hz, the Allan deviation reached a minimum of 50 nA at an averaging time of 500 s. Therefore, the type-A uncertainty ($k = 1$) [26] at an amplitude of 1 A and a frequency of 67 Hz could be derived as 50 nA/A, corresponding to a relative uncertainty of $5 \times 10^{-8}$. For comparison, precision current comparators used in NMIs typically achieve uncertainties of 1 µA/A ($= 1 \times 10^{-6}$) or lower. Thus, the developed device demonstrates a significant advancement, achieving a 20-fold improvement in relative type-A uncertainty.

In addition to high precision, the developed current comparator offers advantages in size and integration. Unlike conventional comparators, it requires neither a detection winding nor massive magnetic core, allowing for a significantly more compact design. Calculations based on a magnetic equivalent circuit model indicate that sensitivity does not degrade with reduced core size. The model suggests that the core can be miniaturized to at least one-tenth of its current dimensions without compromising performance.

In summary, the developed current comparator achieves a relative uncertainty of $5 \times 10^{-8}$, satisfying the requirements for a primary AC current standard. Its compact architecture with strong potential for further miniaturization marks a promising direction for next-generation current metrology.

## Statistical Analysis in DC ratio measurements

To evaluate the versatility of the developed current comparator, we extended the measurements to DC conditions. Accurate operation under DC excitation is critical for ensuring traceability to DC current ratio standards and for verifying the comparator applicability to both AC and DC metrological scenarios. Unlike AC measurements—where the residual magnetic flux component at the excitation frequency can be isolated via DFT—DC measurements require a different approach, as the residual flux cannot be separated from constant background fields using frequency-domain techniques. To overcome this, we applied a square-wave current signal with a duration of 1 s to the ratio windings (top left panel, Fig. 3a). The residual magnetic flux was determined by comparing the NV magnetometer signals during the current-on and current-off states, excluding the transient periods of 500 ms, as indicated by the shaded regions in Fig. 3a. The bottom left panel of Fig. 3a shows the detected magnetic flux when a 1 A DC current was applied. The DC ratio error of the developed current comparator was measured to be 150 nA/A ($1.5 \times 10^{-7}$), based on a residual magnetic flux of approximately 95 pT corresponding to a current difference of 150 nA. This performance is comparable to that of conventional DCCs, which typically achieve ratio errors at the $1 \times 10^{-7}$ level, indicating that the developed device meets the requirements for high-accuracy DC current ratio measurements. Although the error does not yet reach the $10^{-9}$ level attained by CCCs, it is sufficient for several metrological and industrial calibration applications. However, notably, the observed ratio error is condition-dependent and does not represent the ultimate value of the system. Further improvements in shielding and noise suppression may enable even lower ratio errors, approaching those of established DC standards.

The Allan deviation of the DC measurements is shown in Fig. 3b. As in the AC case, the Allan deviation decreased with a $t^{-1/2}$ time dependence, confirming that uncorrelated random noise was dominant. The minimum Allan deviation reached approximately 30 pT, corresponding to a detectable current difference of 50 nA. This yields a type-A uncertainty of 50 nA/A at 1 A input current, corresponding to a relative uncertainty of $5 \times 10^{-8}$. This value is significantly lower than those reported for other NV-diamond-based sensors operating under ambient conditions, which typically exceed 200 µA[27-29].

While CCCs offer sub-nanoampere resolution, they require complex, cryogenic infrastructure. In contrast, the present system achieves nanoampere-level resolution in a compact, room-temperature setup. However, due to low-frequency noise under static conditions (Fig. 1e), achieving this resolution required an averaging time approximately 40 times longer than that needed for AC measurements.

In summary, the DC ratio measurements yielded a ratio error of 150 nA/A and an Allan deviation of 50 nA, exceeding the performance of previously reported NV-based sensors under ambient conditions. These results confirm the suitability of the developed current comparator for precision current ratio metrology in both AC and DC regimes. Future improvements in low-frequency noise suppression and environmental shielding are expected to further enhance its performance and measurement efficiency.

## Conclusions

We have demonstrated a compact, integrated current comparator capable of performing both AC and DC current ratio measurements with $10^{-8}$ accuracy under ambient conditions. By employing a NV diamond-based magnetometer, the system achieves a minimum detectable current of 50 nA, as verified through Allan deviation analysis in both AC and DC regimes. Unlike conventional current comparators—which typically rely on distinct architectures for AC and DC measurements, often requiring bulky windings or cryogenic environments—our approach offers a unified, cryogenics-free solution for high-sensitivity current ratio metrology.

This work establishes a foundation for a simplified and unified current ratio calibration platform compatible with emerging power technologies. While the proposed current comparator demonstrates high precision, some limitations remain in terms of measurement time, frequency range, and current ratio error magnitude. Future efforts will focus on suppressing low-frequency noise, implementing active current injection techniques to reduce frequency-dependent ratio errors, and extending the operating frequency range beyond 300 Hz.

These advancements will not only enhance the performance of current ratio measurements in electric power standards but will also expand the applicability of the system to precision DC resistance bridges and fundamental quantum electrical standards. Ultimately, this work paves the way for compact, cryogenics-free instrumentation capable of supporting a broad range of metrological and industrial applications.

## Acknowledgements

We greatly acknowledge S. Okubo of the optical frequency measurement group at AIST for helpful discussions on Allan deviation. This work was supported by MEXT Quantum Leap Flagship Program (MEXT Q-LEAP) Grant Number JPMXS0118067395, the CSTI, the Cross-ministerial Strategic Innovation Promotion Program (SIP), "Promoting the application of advanced quantum technology platforms to social issues," JST, the Adopting Sustainable Partnerships for Innovative Research Ecosystem (ASPIRE), "Diamond spin qubits for quantum application (DIAMONDQTECH)," and the CSTI, the programs for Bridging the gap between R&d and the IDeal society (society 5.0) and Generating Economic and social value (BRIDGE).

## Author contributions

Y.A., T.I., N.-H.K. and M.H. jointly conceived and led the project. H.M. fabricated the NV-diamond magnetometer and current comparator (CC) devices with the help of Y.K., T.I., H.K., N.S., T.Y., C.U., Y.A., Y.H., and M. Hatano. H.A., S.O., and T.O. conducted electron irradiation of the diamond to generate NV centers in the diamond used for the magnetometer. H.M. conducted the precision measurements and analyzed the data with support from Y.K. and Y. Hatano. H.M., and Y.A. wrote the initial draft. All authors discussed the results and contributed to the final manuscript.

## Competing interests

The authors declare no competing interests